\documentclass[preprint,12pt]{elsart}

\usepackage{graphicx}
\usepackage{epsfig}

\usepackage{amssymb}

\journal{Physica A}

\begin{document}

\begin{frontmatter}

\title{Generalizing Planck's distribution by using the Carati-Galgani model of molecular collisions}

\author{Rubens D. B. Carvalho}
\address{Instituto de Fisica, Universidade Federal do Rio Grande do
Sul, 91501-970, Porto Alegre-RS, Brazil}

\author{Andre M. C. Souza}
\address{Departamento de Fisica, Universidade Federal de
Sergipe, 49100-000 Sao Cristovao-SE, Brazil
\\ National Institute of Science and Technology for
Complex Systems, Rua Xavier Sigaud 150, 22290-180 Rio de Janeiro-RJ,
Brazil } \ead{amcsouza@ufs.br}

\date{\today}

\begin{abstract}
Classical systems of coupled harmonic oscillators are studied using
the Carati-Galgani model. We investigate the consequences for
Einstein's conjecture by considering that the exchanges of energy,
in molecular collisions, follows the L\'evy type statistics. We
develop a generalization of Planck's distribution admitting that
there are analogous relations in the equilibrium quantum statistical
mechanics of the relations found using the nonequilibrium classical
statistical mechanics approach. The generalization of Planck's law
based on the nonextensive statistical mechanics formalism is
compatible with the our analysis.
\end{abstract}


\end{frontmatter}

\section{Introduction}

It was observed that for blackbody radiation\cite{planck} the mean
energy $U$ of stationary electromagnetic waves is a function of
frequency $\nu$ of these waves: at temperature $T$ we can write
\begin{equation}
U(\nu \rightarrow 0) \rightarrow k_B T \;\;\;\; and \;\;\;\; U(\nu
\rightarrow \infty) \rightarrow 0 ,
\end{equation}
where $k_B$ is the Boltzmann constant. This contradicts the energy
equipartition law of classical statistical mechanics in that $U$
is frequency independent. The important contribution of Planck came
when he discovered that one could obtain the experimental result of Eq. (1)
by treating the energy of the stationary electromagnetic waves as a
discrete spectra. Planck's law opened the way for quantum
mechanics.

In such a way, Einstein \cite{einstein} gave an important contribution by
noticing that in the canonical ensemble the fluctuation of energy of the
blackbody radiation is compatible with Planck's law when it is,
in terms of its variance $\sigma_E$, expressed in terms of the mean
energy $U$ through the relations
\begin{equation}
\frac{dU}{d\beta}=-(\sigma_E)^2  \label{ein1}
\end{equation}
and
\begin{equation}
(\sigma_E)^2=\epsilon U + \frac{U^2}{N} \;  \label{ein2}
\end{equation}
where $\beta=1/ (k_B T)$, $\epsilon=h\nu$ ($h$ is Planck's
constant) and $N$ is the number of quantum particles. Einstein imagined
the first relation as being exactly a type of general thermodynamic
relation, while the second must have a dynamical character, being, in
principle, deducible from a microscopic dynamics.

Recently, Carati and Galgani \cite{galgani,gal2} have
investigated classical systems of coupled harmonic oscillators with
the procedure of Jeans-Landau-Teller dynamic of molecular collisions
\cite{JLT,JLT2}. The principal results of Carati-Galgani are that:
i) relaxation times to equilibrium are nonuniform and depend on the
internal degrees of freedom of the considered system; ii) Situations
of nonequilibrium statistical mechanics very far from equilibrium
are described, in a first order approximation, by the analogous relations found
in quantum equilibrium statistical mechanics; iii) as
conjectured by Einstein, Eq. (\ref{ein2}) is a consequence of the
dynamics.

It is important to remark that Planck's law considers a real
equilibrium state of the quantum statistical mechanics, while within
the framework of Carati-Galgani, a similar relation to Planck's
law was obtained using nonequilibrium classical statistical
mechanics.

The purpose of the present work is to study classical systems of
coupled harmonic oscillators along the line of Carati-Galgani.
We studied three different aspects of this framework. First, we
analyzed the consequence for Einstein's conjecture by assuming
that the exchanges of energy, in molecular collisions, follows the
L\'evy type statistic for high enough frequencies. The L\'evy
statistics for the energy exchanges was recently obtained by Carati
et al. \cite{carprl}, considering the dynamics over a finite time in
the Jeans-Landau-Teller model. Secondly, admitting that there are
analogous relations in the equilibrium quantum statistical mechanics
of the relations found here using the nonequilibrium classical
statistical mechanics approach, we will explore some possible
consequences in quantum radiation theory.  Third, we
show that the generalization of Planck's law \cite{euct} within the
framework of the recently introduced superstatistic formalism
\cite{beckcohen,tsza} is compatible with the our analysis of Einstein's conjecture.

 The organization of this paper is as
follows. The Carati-Galgani model is briefly explained in Sec. II.
Our generalization is presented in Sec. III. Discussions about the
consequences of our results are presented in Sec. IV. Concluding
remarks are given in Sec. V.

\section{The Carati-Galgani model}

A good level of interest has been focused on the problem of estimate
energy exchanges between vibrational and translational degrees of
freedom in the dynamic of molecular collisions. One of the simplest
and main model used to study this problem was published by Carati
and Galgani \cite{galgani,gal2}. It is based on the
Jeans-Landau-Teller approach \cite{JLT}. A possible example of a
physical system that reproduce this formalism follows.

Let us imagine a one-dimensional system involving two
particles $P$ and $Q$ in a line. $P$ is attracted by a linear spring
to the fixed origin through a harmonic oscillator potential. $Q$
interacts with $P$ through an analytical potential. In a first
approximation, the variation of energy $\delta e$ in a simple
collision between $P$ and $Q$ for the oscillator $P$ of frequency
$\nu$ is given by the addition of a drift term and others dependent
on the fluctuation of the oscillator's initial phase. It is easy to
show \cite{galgani} that after $t$ collisions an oscillator has
energy
\begin{equation}
\epsilon_{t}= \epsilon_{0} + t\eta^{2} + 2\eta \sum_{j=1}^{t}
\sqrt{e_{j-1}} \cos (\varphi_{j-1}) . \label{energy}
\end{equation}
$\epsilon_{0}$ is the oscillator's initial energy that is considered
to be very small in order that the motion of $P$ and $Q$ be
decoupled. $\varphi_j$ is the phase of the oscillator in the $j$th
collision and $\eta$ is evaluated as a function of $\nu$, which is
known to be decreasing exponentially with $\nu$ (for an analysis of
$\eta$ see \cite{JLT2}).

Let us consider a assembly of oscillators, averaging over the phases
$\varphi_j$, in order that
\begin{equation}
\langle\epsilon_{t}\rangle= \langle\epsilon_{0}\rangle + t\eta^{2} +
2\eta \sum_{j=1}^{t} \langle\sqrt{e_{j-1}} \cos
(\varphi_{j-1})\rangle . \label{enmean1}
\end{equation}
where $\langle...\rangle$ is the phase average. Considering the
phases to be independent and uniformly distributed, for large $t$, the
average of energy and the variance are, respectively, given by
\begin{equation}
\langle\epsilon_{t}\rangle= \langle\epsilon_{0}\rangle + t\eta^{2}
 \label{enmean2}
\end{equation}
and
\begin{equation}
\sigma^2_{\epsilon_{t}}= 2\epsilon_{0} \langle\mu_{t}\rangle +
\langle\mu_{t}^2\rangle \;. \label{cara0}
\end{equation}
Here $\sigma^2_{\epsilon_{t}}$ and $\langle\mu_{t}\rangle$
correspond to, respectively, the variance and mean of the exchanged
energy $(\epsilon_{t}-\epsilon_{0})$. Observe that the functional
form of the variance is independent of $t$ and $\eta$.

Assuming $N$ independent identical oscillators and considering the
total energy $\langle E_{t} \rangle = \sum_j \langle
\mu_{t}^{j}\rangle$, we immediately obtain for the variance
$\sigma_E^2 = N \sigma_\epsilon^2$ and mean $U=N\langle \mu \rangle$
of the exchanged energy
\begin{equation}
(\sigma_E)^2= 2\varsigma \nu U + \frac{U^2}{N}  \label{carat}
\end{equation}
where $\varsigma$ is the initial action per oscillator \cite{galgani}.
This result represents essentially the application
of the central limit theorem, that arise
from a large number of independent contributions of the oscillators \cite{bill} .

Equation (\ref{carat}) is analogous to Einstein's
relation obtained in the Bose-Einstein quantum statistical mechanics
theory. Despite the fact that this relation has been obtained considering the
classical statistical mechanics very far from equilibrium, the
approach recovers Einstein's conjecture, i.e., proving it to be a
consequence of the dynamics.

\section{The generalization of the Einstein's conjecture}

Recently, Carati et al. \cite{carprl} showed that there
are systems which present qualitative differences between the
statistics obtained using the equilibrium distribution and that
obtained through the dynamics over a finite time. Considering the
dynamics over a finite time of the Jeans-Landau-Teller model, they
obtained that the exchanges of energy follow the L\'evy type
statistics for high enough frequencies. An immediate consequence of
this result, based on the L\'evy flight random walk \cite{vis}, is that
$\langle\mu_{t}^2\rangle$ is infinite.


We notice that in the Carati-Galgani model it is assumed that the
oscillators have low energies and the collisions are separated
from each other so that the oscillator phases are completely
independent. Here, we must be able to answer what happens if the $\langle\mu_{t}^2\rangle$ is not finite? 
In this way, we can no longer write Eq. (\ref{cara0}) and, as consequence, Eq.
(\ref{carat}) is not valid either. 

It is well known that for $\langle\mu_{t}^2\rangle \rightarrow
\infty$ the theoretical basis is the Lindeberg-L\'evy central limit
theorem \cite{gne}. Since we are
considering the case of low energies for the oscillators, 
using the same steps as were used to obtain Eq. (\ref{cara0}),
it is straightforwardly verified that
\begin{equation}
\sigma^2_{\epsilon}= 2\epsilon_{0} \langle \mu\rangle +
\varepsilon^{(2-\alpha)}\langle \mu^{\alpha} \rangle \;\;\;\;\;\; 1 \leq \alpha < 2
\;, \label{levy}
\end{equation}
where $\varepsilon$ is a constant that has dimension of energy and
is suitably chosen by dimensional requirements.

Finally, for $N$ independent identical oscillators, we find
\begin{equation}
\sigma^2_E = 2\varsigma \nu U + \varepsilon^{(2-\alpha)}
\frac{U^\alpha}{N^{\alpha -1}} \label{our}
\end{equation}
which generalizes the dynamic proposal of Einstein and will be
our hypothesis to get the generalizations of Einstein's conjecture and Planck's law.

\section{Possible relations in equilibrium statistical mechanics}
Let us now explore the possibility of there existing analogous
relations in the equilibrium quantum statistical mechanics of the
relations found here using the nonequilibrium classical statistical
mechanics approach.

We start the process using Eq. (\ref{ein1}) in conformity to Einstein's vision.
In this case, using Eq. (\ref{our}) in Eq. (\ref{ein1}) we have
\begin{equation}
-\frac{dU}{d\beta}= 2\varsigma \nu U + \varepsilon^{(2-\alpha)}
\frac{U^\alpha}{N^{\alpha -1}}. \label{our2}
\end{equation}
It is straightforward to find the energy distribution
\begin{equation}
U = N \{ \frac{2\varsigma
\nu}{\varepsilon^{(2-\alpha)}[e^{(\alpha -1)2\varsigma \nu\beta}-1]}
\}^{\frac{1}{\alpha -1}} \; , \label{plg}
\end{equation}
in order that we can consistently rewrite Eq. (\ref{plg}) as
\begin{equation}
\tilde{U} = \frac{ H \nu}{(e^{ \beta H \nu /d}-1 )^{d}}  \; .
\label{pld}
\end{equation}
We have introduced $\tilde{U}\equiv \frac{Uz^{d-1}}{N}$, $H\equiv
2\varsigma$ and  $d \equiv \frac{1}{\alpha -1}$ with $1 \leq d < \infty$.
We assumed $\varepsilon \propto H \nu$ that leads to the expression
$\varepsilon = z H \nu$ with $z$ a pure dimensionless number. For
$d=1$, $H$ coincides with Planck's constant $h$, and the $U$
associated with Bose-Einstein quantum statistic is reobtained.

\begin{figure}
\includegraphics[width=110mm]{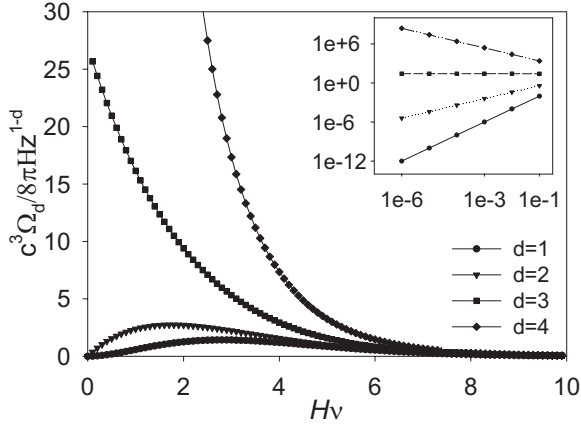}
\caption{The density of energy per unit volume versus energy for $d=1,2,3,4$ and $\beta=1$.
(Inset shows a closeup of part of the graph for $\nu \rightarrow 0$)}
\label{fig1}
\end{figure}

\begin{figure}
\includegraphics[width=110mm]{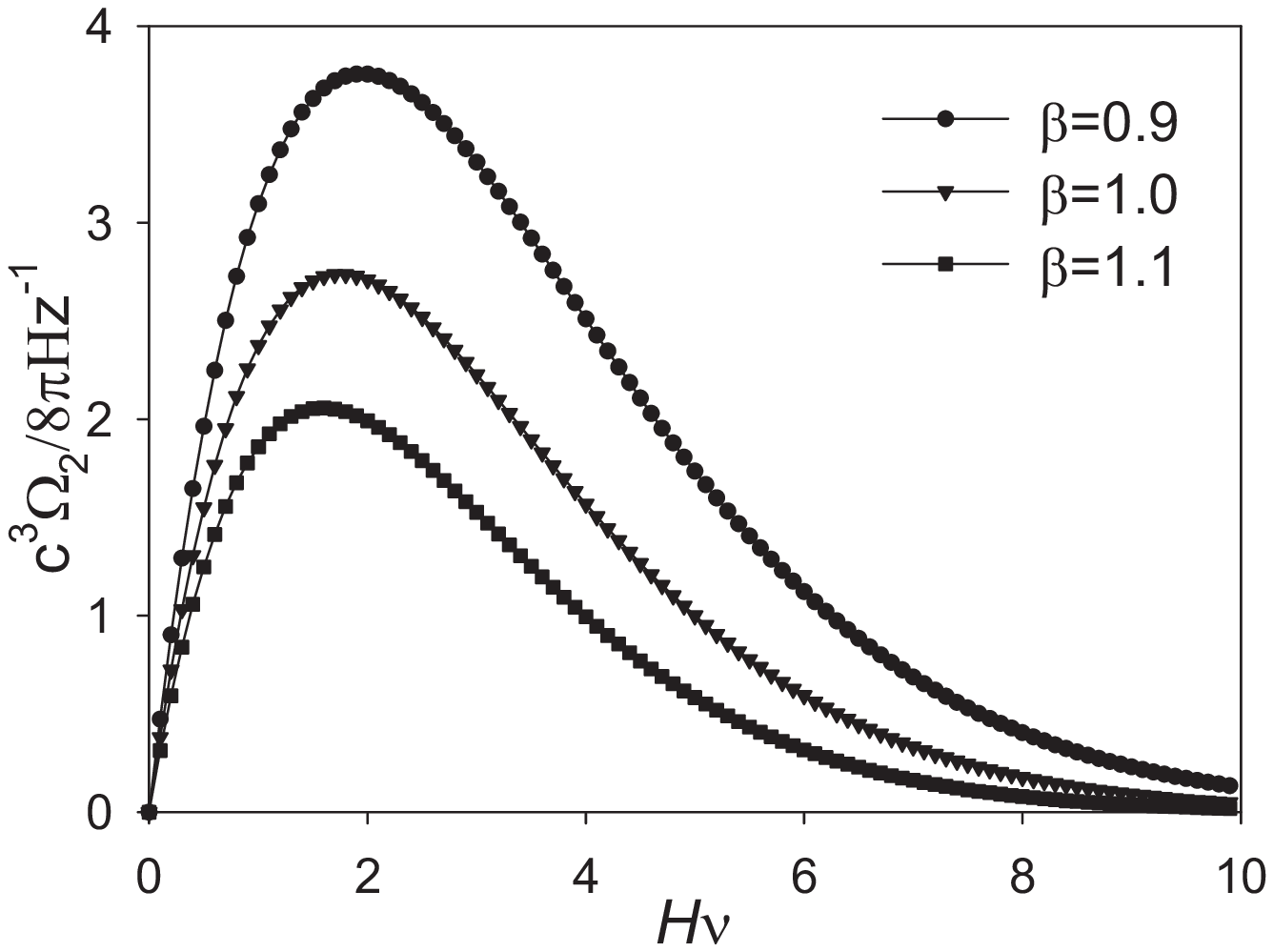}
\caption{The density of energy per unit volume versus energy for $\beta=0.9,1.0,1.1$ and $d=2$.}
\label{fig2}
\end{figure}

\begin{figure}
\includegraphics[width=110mm]{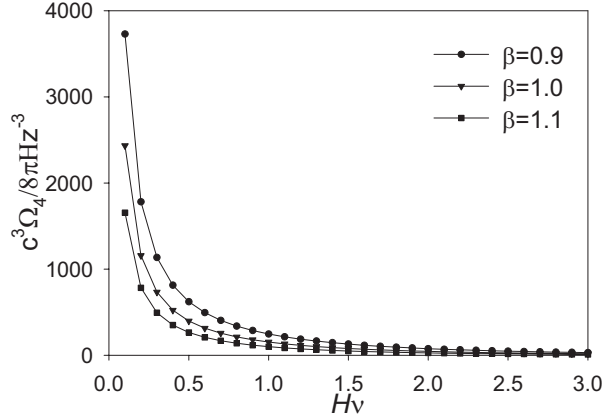}
\caption{The density of energy per unit volume versus energy for $\beta=0.9,1.0,1.1$ and $d=4$.}
\label{fig3}
\end{figure}

In the same way as the Planck blackbody radiation has its origin
in the Bose-Einstein statistics, here obtained using $d=1$, we can
take other values of $d$, so that we can obtain new
physical systems such that Eq. (\ref{pld}) may be applied. By
following along standard lines, after straightforward calculations, the
density of energy per unit volume is obtained as
\begin{equation}
 \Omega_d (\nu)= g(\nu) U =
\frac{8\pi H \nu^{3}}{z^{d-1}c^{3}[e^{\beta H \nu /d}-1]^{d}}
\label{ome}
\end{equation}
where we use the standard density of states of the photons ($g(\nu)=
8 \pi \nu^2/c^3$). Fig. \ref{fig1} presents $\Omega_d (\nu)$ versus
energy $H \nu$ for typical values of $d$ and $\beta=1$. Note that if
$d$ increases then $\Omega_d (\nu)$ increases.
Although $\Omega_3 (0)$ has a finite value for $\nu \rightarrow 0$, it is null for $d < 3$
while it is infinity for $d > 3$ (see inset). The $\Omega_d$ is modified
significantly when $d$ varies.
The $\Omega_d$ are illustrated in Figs. \ref{fig2} and \ref{fig3} for,
respectively, $d=2$ and $d=4$ and various values of $\beta$. For $d$ fix, we note
small changes for $\Omega_d$ when $\beta$ varies.

In the $H \nu \beta \ll 1$ regime
\begin{equation}
 \Omega_d (\nu) \sim
\frac{8 \pi \nu^{3-d}}{(zH)^{d-1}c^{3}} (d K_B T)^{d} \;  \label{rj}
\end{equation}
which generalizes the Rayleigh-Jeans law. The generalization of Wien's law
is obtained in the $H \nu \beta \gg 1$ region
\begin{equation}
 \Omega_d (\nu) \sim
\frac{8 \pi H \nu^{3}}{z^{d-1}c^{3}} e^{-\frac{H \nu}{ K_B T} } \; ,
\label{wie}
\end{equation}
and the Stefan-Boltzmann law for the total emitted power per unit
surface is attained from the usual one, given by $j=\sigma_d
T^{4}$, where the d-dependence is present only in the prefactor
$\sigma_d$.

\section{Souza-Tsallis generalization from the Einstein's conjecture}
In fact, one of the characteristics of Eq. (\ref{pld}) is that it
reproduces the results of Souza and Tsallis \cite{euct}. Observe
that for $\tilde{U} = H \nu \bar{n}$ we can obtain the
particle-number distribution
\begin{equation}
\bar{n} = \frac{1}{(e^{ \beta H \nu /d}-1 )^{d}}  \; . \label{nud}
\end{equation}
Adopting the formal theory of statistical mechanics, it is
straightforward to find the variance of an ensemble of $n$ particles
as being
\begin{equation}
\sigma^{2}_{n}= \bar{n} + \bar{n}^{\alpha} \label{nsi}
\end{equation}
which generalizes the bosonic Landau's relation \cite{landau}.

Ref. \cite{euct} obtained the Eqs. (\ref{nud}) and (\ref{nsi})
heuristically based on the nonextensive statistical mechanics
\cite{tsa} and Beck-Cohen superstatistics \cite{beckcohen,tsza}.
They were based on the discussion of the differential equation $dy/dx=-a
_{1}y-(a _{q}-a_{1})\,y^{\alpha}$ (with $y(0)=1$), whose particular case
$\alpha =2$ corresponds to Bose-Einstein statistics. Our formalism
describes the Souza-Tsallis generalization from the Einstein's
dynamical fluctuation viewpoint.

Finally, we see that Eq. (\ref{nud}) can be written as
\begin{equation}
\bar{n} =  \sum_{i=0}^\infty g(i,d)\, e^{-\beta E_i} \,,
 \end{equation}
where we obtain the energy spectrum
\begin{equation}
E_i (d) \equiv \frac{H \nu}{d} (i+d) ,
\end{equation}
and its degeneracy
\begin{equation}
g(i,d) \equiv \frac{\Gamma \left( i+d \right) }{\Gamma \left( d
\right) \Gamma \left( i+1\right) }  \,,
\end{equation}
$\Gamma (x)$ being the gamma function. The spectrum is made of
equidistant levels, like that of the quantum harmonic oscillator,
with a constant energy difference between successive levels equal to
$H\nu /d$.

\section{Conclusions}
Summarizing, we introduced a generalization of Einstein's conjecture
considering an approach along the line of the Carati-Galgani model
of molecular collisions. We obtained Planck's distribution
compatible with the Beck-Cohen superstatistics. Furthermore,
establishing a metaequilibrium very far from equilibrium for the
system studied, and admitting that there are analogous relations in
the equilibrium quantum mechanics, we generalized Planck's law.

For instance, the effects in the quantum radiation theory were
obtained theoretically, as a first order approximation, of a
classical system very far from equilibrium \cite{gal2}. We emphasize
that we are working with classical models of molecular collisions
and not quantum models for blackbody radiation. Understanding the
relationship between these two points of view is an open problem.

\section*{ACKNOWLEDGMENTS}
This work was supported by CNPq (Brazilian Agency).

\bibliographystyle{elsarticle-num}

\end{document}